\documentstyle[aps]{revtex}
\begin{document}
\title{How to realize quantum superluminal communication?}
\author{Gao shan}
\address{Institute of Quantum Mechanics, 11-10, NO.10 Building,
YueTan XiJie DongLi, XiCheng District, Beijing 100045, P.R.China.
Tel: 86 010 82882274  Fax: 86 010 82882019  Email:
gaoshan.iqm@263.net}
\date{June 14, 1999}
\maketitle

\vskip 0.5cm

\begin{abstract}
We deeply analyze the possibility to achieve quantum superluminal
communication beyond the domain of special relativity and present
quantum theory, and show that when using the conscious object as
one part of the measuring device, quantum superluminal
communication may be a natural thing.
\end{abstract}

\section{Introduction}

After having shown quantum superluminal communication must exist
in our world\cite{Gao4}, we will further study the possibility to
realize such superluminal communication, this is undoubtedly a
formidable task within the scope of our present knowledge about
Nature, since the two foundation stones of modern
physics---special relativity and quantum mechanics all reject
superluminal communication, thus in order to achieve quantum
superluminal communication, we must first revise them.

\section{Revising special relativity}
In order to admit quantum superluminal communication, special
relativity needs to be changed only a little, and this does not
limit its applicability in its previous territory at all, in fact,
we only need to limit the scope of "natural phenomena" in the
principle of relativity, which is the first assumption of special
relativity, namely the natural phenomena satisfying the principle
of relativity will no longer involve all natural phenomena, and
the quantum nonlocal influence is just such an
exception\cite{Gao3}, we call the new principle quantum relativity
principle.

In fact, there exists nothing compelling in both theoretical
considerations and experimental confirmations to require the
validity of the principle of relativity for all natural phenomena,
just as Einstein, the founder of special relativity, demonstrated
himself\cite{Einstein1}, "in view of the more recent development
of electrodynamics and optics, it became more and more evident
that classical mechanics affords an insufficient foundation for
the physical description of all natural phenomena. At this
juncture the question of the validity of the principle of
relativity became ripe for discussion, and it did not appear
impossible that the answer to this question might be in the
negative." Indeed, his worries become true when considering the
quantum nonlocal influence in quantum theory.

On the other hand, although we have demonstrated that the quantum
nonlocal influence rejects the relativity principle owing to the
resulting causal loop\cite{Gao3}, the deeper reasons need to be
given, as Einstein denoted\cite{Einstein1}, the relativity
principle originates from classical mechanics, the precondition of
its validity will be "all natural phenomena were capable of
representation with the help of classical mechanics", while
quantum phenomena are evidently not such natural phenomena, and
classical mechanics can no longer affords an sufficient foundation
for the physical description of such phenomena either, concretely
speaking, the relativity principle will hold true for the
continuous motion of the real objects including particles and
fields, while for the quantum nonlocal influence in quantum
space-like measurement, no real objects are transmitted in the
process, and the process is also essentially discontinuous, thus
there does not exist any real objects in continuous motion for the
principle to apply in such process, and the earth beneath the feet
indeed disappears, then it is by no means a surprising fact that
the quantum nonlocal influence does reject the relativity
principle\cite{Gao3}.

Indeed, if one principle is valid for all natural phenomena, it
will be too absolute to be true, the original relativity principle
is just such a principle; on the other hand, its founder Einstein
also ignored one subtle possibility, namely the invalidity of the
relativity principle for some natural phenomena, say the quantum
nonlocal influence, will not influence its validity for other
natural phenomena, say classical phenomena.

At last, even though special relativity is revised so as to permit
the existence of quantum superluminal communication, it provides
nothing helpful for realizing such superluminal communication,
since the origin lies in the quantum nonlocal influence itself,
thus we must turn to the quantum theory.

\section{Revising quantum mechanics}
According to the demonstrations about the existence of quantum
superluminal communication\cite{Gao4}, we know that in order to
find the preferred Lorentz frame, which existence is predicted in
theory, the distinguishability of nonorthogonal single states is
required, but present quantum theory uncompromisingly rejects this
requirement\cite{Eber}, thus we need to revise present quantum
theory.

First, as to the normal linear evolution equation of the wave
function in present quantum theory, even though it has been
confirmed very precisely, we can not exclude the possibility of
its inaccuracy yet, and there may exist some kind of deterministic
nonlinear correction, which will result in the different
predictions from present quantum
mechanics\cite{Weinberg,Gisin,Polchinski}, then quantum
superluminal communication can be achieved in such revised quantum
theory.

Secondly, as to the evolution of the wave function during quantum
measurement, present quantum theory is by no means a complete
theory, and the projection postulate, from which the quantum
nonlocal influence appears, is just a makeshift, while the
concrete dynamical process of the projection, where the
availability of quantum superluminal communication may hide, is
undoubtedly one of the most important unsettled problems in
quantum theory, and the resulting revised theories are deeply
studied
recently\cite{Dio89,Gao,Ghir86,Ghir90,Pea86,Pea89,Pen86,Pen96}, in
which the linear evolution equation of the wave function is
replaced by stochastic linear or nonlinear equation; on the other
hand, there exists no essential reason to prevent the revised
quantum theory including dynamical collapse process from
permitting the distinguishability of nonorthogonal single states,
on the contrary, this kind of distinguishability will disclose the
mysterious veiling of the wave function, say endow with reality to
it, and eventually settle the notorious interpretation problem of
quantum mechanics.

\section{How to realize quantum superluminal communication?}
Since no essential reason and experimental evidence can be found
to revise the normal linear evolution equation of the wave
function by some deterministic nonlinear correction, we will
mainly analyze the possibility to achieve quantum superluminal
communication by use of the revised quantum theory including
dynamical collapse process.

First, even though present quantum theory, especially its
projection postulate, is surely incomplete, but in case of its
correctness it still imposes strong limitations for the
availability of quantum superluminal communication, concretely
speaking, once the quantum measurement process is completed,
projection postulate will take effect, then little room is left
for the possibility of quantum superluminal communication, thus
the opportunity only exists in the quantum measurement process.

Secondly, since for any quantum measurement the measurement
results undoubtedly need to be obtained by the observer, say our
human being, or other conscious object, who wants to carry out
such measurement, then the above opportunity requires that the
state of the conscious object must be entangled with the measured
state before the completion of the quantum measurement process,
two essential advantages to resort to conscious are that, on the
one hand, the conscious object will be the last identifier in the
measurement process and its influence is inescapable before the
dynamical collapse process is completed, on the other hand, only
the conscious object may identify the intermediate process and
state before the dynamical collapse process is completed owing to
his self-conscious, and takes a different action from the action
corresponding to one of the states in the superposition state.

Since the distinguishability of nonorthogonal single states is
required to achieve quantum superluminal communication, here we
assume what need to be measured and differentiated are the
following nonorthogonal single states $\psi_{1}+\psi_{2}$ and
$\psi_{1}$, and the entangled state of the whole system including
measured system, measuring device and conscious object is
respectively
$\psi_{1}\varphi_{1}\chi_{1}+\psi_{2}\varphi_{2}\chi_{2}$and
$\psi_{1}\varphi_{1}\chi_{1}$, where $\varphi_{1}$ and
$\varphi_{2}$ are the states of the measuring device, $\chi_{1}$
and $\chi_{2}$ are the conscious state of the conscious object,
then it is evident that, in order to distinguish the above
nonorthogonal single states, the only reasonable condition is the
conscious time for the definite state
$\psi_{1}\varphi_{1}\chi_{1}$ is shorter than the collapse time
for the superposition state
$\psi_{1}\varphi_{1}\chi_{1}+\psi_{2}\varphi_{2}\chi_{2}$, and the
time difference is long enough for the conscious object to
identify\footnote{This condition may result in the conclusion that
conscious can not be explained by present ( quantum ) physical
theory.}.
%Ghirardi's suggestions?!!!

In the following, we will mainly demonstrate that this condition
is not irrational, and can be satisfied in essence, first, we can
assume the conscious time of the conscious object in Nature is
generally independent of the collapse process of the observed
state, since during the formation and evolution of conscious the
input states will be always classical definite states coming from
the outside classical world, the conscious object can only be
trained to adapt to these classical definite states, and there
exists no collapse process for the observed classical state at
all, furthermore, with the natural selection the conscious time
will turn shorter and shorter, while the universal collapse time
formula is not changed, then it is reasonable that for some kind
of conscious object the conscious time for the definite state
$\psi_{1}\varphi_{1}\chi_{1}$ is shorter than the collapse time of
the superposition state
$\psi_{1}\varphi_{1}\chi_{1}+\psi_{2}\varphi_{2}\chi_{2}$, and the
time difference is long enough for the conscious object to
identify, thus even if our human being can not satisfy this
condition, other conscious objects may satisfy this condition.

On the other hand, if the above condition can not be satisfied in
essence, then we must accept the following bizarre conclusion,
namely the concrete collapse theory will limit both the mass, size
of the conscious part and conscious time of any conscious object,
this means that in case of a certain conscious time the mass and
size of the conscious part of any conscious object can not be
smaller than the minimal finite values, and in case of a certain
mass and size the conscious time of any conscious object can not
be shorter than the corresponding collapse time, these
requirements are evidently irrational, since the above properties
of any conscious object all originate from the natural selection
in the environment of classical world, and even if the natural
selection relating to the collapse process does take effect, then
it is also more reasonable for the conscious object to be able to
use quantum superluminal communication, since undoubtedly it will
be useful for his existence and evolution.

\section{A feasible experiment}
In this section, we will present a feasible experiment to confirm
the above possibility of quantum superluminal communication.

The experiment is based on the fact that the visual perceptual
apparatus of many creatures including our human being possesses an
extreme sensitivity, and as we know, for hoptoad only one photon
can trigger a definite visual perception, for our human being the
number is about seven\cite{Penrose}.

We first consider a simple case in which a bunch of 10 photons
coming from a region A propagates towards the eye of a human
observer\cite{Ghir}, which has been analyzed by Ghirardi from a
different point of view, the photons hit the retina of the
observer and trigger the definite perception ``a luminous spot at
A''; in a similar way, the photons coming from a region B,
spatially separated and perceptively distinct from A, will trigger
the definite perception of the observer ``a luminous spot at B''.

Now, we consider a different situation, in which a superposition
state of the above two states of the photons is prepared and kept
long enough for the experimental aim using present quantum optics
technology, and again the photons propagate towards the eye of the
same human observer, hit his retina and trigger his visual
perception, then according to the above analysis the observer may
have a different perception from any one of the above perceptions.

Certainly, due to the brevity of the perceptual process, it may be
very difficult for the observe to perceive the difference of the
above two situations, but as we think, it is by no means
impossible some conscious object in nature.

On the other hand, the human observer may be replaced by a trained
hoptoad, and the above experiment may contain only one photon,
which can be easily realized presently, then we can observe the
different reactions of the hoptoad for the above two situations,
according to our analysis, for the superposition state situation
the hoptoad may take a different reaction from any one of the
reactions corresponding to the definite states in the
superposition state.

\section{Conclusions}
We show that special relativity and present quantum theory can be
revised to permit quantum superluminal communication, and when
using the conscious object as one part of the measuring device,
quantum superluminal communication may be a natural thing.

\vskip 1cm \noindent Acknowledgments \vskip .5cm Thanks for
helpful discussions with Dr S.X.Yu ( Institute Of Theoretical
Physics, Academia Sinica ).

\end{document}